\documentclass[sigconf]{acmart}
\usepackage{threeparttable}
\usepackage{booktabs}
\usepackage{graphicx}
\usepackage{multirow}

\AtBeginDocument{%
  \providecommand\BibTeX{{%
    \normalfont B\kern-0.5em{\scshape i\kern-0.25em b}\kern-0.8em\TeX}}}

\setcopyright{acmcopyright}
\copyrightyear{2022}
\acmYear{2022}
\acmDOI{}

\acmConference[WebConf '22]{GLB 2022@The ACM Web Conference 2022.}{April 25--29, 2022}{Lyons}
\acmBooktitle{GLB 2022@WebConf '22: The ACM Web Conference 2022, April 25--29, 2022, Lyons}
\acmPrice{15.00}
\acmISBN{}
\usepackage{booktabs}
\usepackage{multirow}
\usepackage{graphicx}
\usepackage{subfigure}

\usepackage{algorithm}
\usepackage{algorithmic}

\begin{document}

\title{TeleGraph: A Benchmark Dataset for Hierarchical Link Prediction }

\setlength{\abovedisplayskip}{-0.1pt}
\setlength{\belowdisplayskip}{-0.1pt}
\setlength{\abovecaptionskip}{-0.1pt}
\setlength{\belowcaptionskip}{-0.1pt}
\setlength{\floatsep}{-0.1pt}
\setlength{\textfloatsep}{-0.1pt}

\author{Min Zhou}
\affiliation{
  \institution{Huawei Noah's Ark Lab}
  \city{Shenzhen}
  \country{China}
}
\email{zhoumin27@huawei.com}

\author{Bisheng Li}
\affiliation{
 \institution{Fudan University}
 \city{Shanghai}
 \country{China}
}
\email{bsli20@fudan.edu.cn}

\author{Menglin Yang}
\affiliation{
 \institution{The Chinese University of Hong Kong}
 \city{Hong Kong}
 \country{China}
}
\email{mlyang@cse.cuhk.edu.hk}

\author{Lujia Pan}
\affiliation{
  \institution{Huawei Noah's Ark Lab}
  \city{Shenzhen}
  \country{China}
}
\email{panlujia@huawei.com}


    
  





\renewcommand{\shortauthors}{Zhou et al.}

\begin{abstract}

Link prediction is a key problem for network-structured data, attracting considerable research efforts owing to its diverse applications. 
The current link prediction methods focus on general networks and are overly dependent on either the closed triangular structure
of networks or node attributes. Their performance on sparse or highly hierarchical networks has not been well studied.  On the other hand,   the available tree-like benchmark datasets are either simulated, with limited node information, or small in scale. To bridge this gap, we present a new benchmark dataset \textit{TeleGraph}, a highly sparse and hierarchical telecommunication network associated with rich node attributes, for assessing and fostering the link inference techniques.  Our empirical results suggest that most of the algorithms fail 
to produce satisfactory performance on a nearly tree-like dataset, which calls for special attention when designing or deploying the link prediction algorithm in practice.

\end{abstract}

\maketitle

\section{Introduction}

Link prediction aims to predict whether two nodes in a network are likely to have a link~\cite{liben2007link,yang2021discrete} via the partially available topology or/and node attributes, attracting considerable research efforts owing to its diverse applications. According to the techniques involved, the current link prediction methods can be categorized into three classes: heuristic methods, embedding methods, and Graph Neural Network (GNN)-based methods. Heuristic methods infer the likelihood of links via handcrafted similarity measures regarding the structure information ~\cite{liben2007link,barabasi1999emergence}. Embedding methods learn free-parameter node embeddings in a transductive manner to infer the probability if a node pair being connected~\cite{node2vec-kdd2016,koren2009matrix}. GNN-based methods formulate the link prediction as the binary classification problem in which the  explicit node feature could be incorporated~\cite{kipf2016variational,zhang2018link,yun2021neo}. These techniques are typically evaluated on a limited number of regular benchmark datasets such as collaboration network~\cite{wang2020microsoft} or citation networks ~\cite{huang2021scaling}. 
However, many real-world scenarios, such as protein interaction networks~\cite{xu2018powerful}, knowledge graphs~\cite{tifrea2018poincare}, disease spreading network~\cite{chami2019hyperbolic}, transport networks~\cite{ballantyne2018fault}, and social networks~\cite{bearman2004chains}, are usually sparse or hierarchical in nature, which calls for special attentions ~\cite{shang2019link,bearman2004chains,yang2022hyperbolic}.  On the other hand, the current tree-like benchmark datasets are either simulated, small in scale, or with limited node attributes, which incurs substantial affect on the progress of related studies.

\begin{figure}[!tp]
    \centering
    \includegraphics[width=0.85\linewidth]{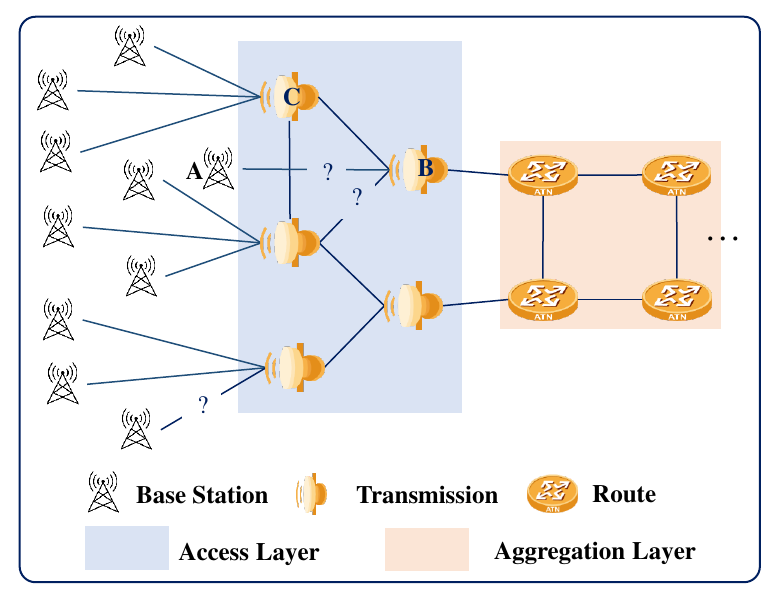}
    \caption{An illustration of the access network which is the most basic facility in the telecommunication network.  It is responsible for connecting subscribers to the immediate service provider via base stations, transmission, and route.  The overall topology is tree-like with a distinct hierarchical layout, but some local areas are grid or circular. Recovering the missing topology between the devices enables the efficient fault management ~\cite{fournier2020discovering}, which is of importance to ensure the stability and reliability of the network.  }
    \label{fig:layout}
    \vspace{5pt}
\end{figure}

To bridge the gap and further foster the link prediction research, we publicly release \textit{TeleGraph}: a medium-sized (41,143 nodes) undirected heterogeneous network (3 classes) with multiple informative node attributes (240 types). Telecommunication networks are keys to supporting personal communications as well as those of businesses and other organizations in the modern era. To ensure the stability and reliability of the large telecommunication network, a crucial task is efficient fault management, which requires the precise device connections enabling the engineers to analyze, locate, and recover the faults~\cite{fournier2020discovering}.  An illustration of the access layer of a telecommunication network is given in Figure~\ref{fig:layout}, where the overall topology is tree-like but some local areas are grid or circular, and some connections between devices are missing.  As a real-world dataset associated with complex topology and multiple node attributes and whose properties are shared by many engineered transport networks~\cite{ballantyne2018fault}, we believe \textit{TeleGraph} creates exciting opportunities for assessing and gestating both link inference techniques and node embedding procedures.

\textbf{Main contributions.} The major results and contributions presented in our work can be summed up as follows:
\begin{itemize}
    \item We release TeleGraph, a highly hierarchical and sparse dataset which can be used to benchmark the link prediction algorithms.
    \item We conduct a descriptive analysis of the dataset and discuss the particular modeling challenges that the dataset poses.
    \item We assess the performance of existing link prediction algorithms regarding to both AUC and AP, provide insights and prescriptive guidance for industrial settings from our observations.
\end{itemize}

\section{Preliminary}

\subsection{Notations and Problem Definition}
Consider an undirected graph $G = (V, E,X)$ with  node set $V$, the edge set $E$, and node features $X$, respectively. We denote $A\in\{0,1\}^{N\times N}$ as the adjacency matrix of $G$, i.e., the $ (i,j)$-th entry in $A$ is 1 if and only if there is an edge in $E$ between $v_i$ and $v_j$ ($v\in V$). Given network structure or/and node features, link prediction methods aim to preserve/learn the similarity measures of the node pair to infer missing links or detect the spurious links.

 According to the techniques involved, the current Link prediction algorithms can be generally classified into three main paradigms: heuristic methods, embedding methods, and GNN-based methods. Heuristic methods compute some heuristic node similarity scores as the likelihood of links~\cite{liben2007link, lu2011link}, which are simple yet effective but not applicable for diverse scenarios.  Embedding methods learn node embedding based on connections between nodes and compute similarity scores~\cite{perozzi2014deepwalk,node2vec-kdd2016}. GNN-based methods infer the existence of links by employing graph neural networks. More specifically, Graph Auto-Encoder(GAE) or Variational GAE~\cite{kipf2016variational} learn node representations through an auto-encoder framework where various GNN architectures have been utilized as the encoders. SEAL~\cite{zhang2018link,zhang2021labeling} reformulated the link prediction task as the subgraph binary classification which shows apparent advantages over GAEs in most scenarios.  In Section ~\ref{sec:exp}, we will carry on extensive experiments to evaluate the performance of diverse link prediction methods.

\subsection{Evaluation Metrics} 
To test the algorithms' performance, the edge set $E$ is randomly divided into the training, validation, and test set. Two standard metrics: area under the receiver operating characteristic curve (AUC) ~\cite{zhou2009predicting} and Average Precision (AP) ~\cite{Zhang2009} are widely applied to evaluate the link prediction measures or algorithms. 

\textbf{(i) AUC:} Given the ranking of the non-observed links, the AUC value can be interpreted as the probability that a randomly chosen missing link  is given a higher score than a randomly chosen nonexistent link. At each time we randomly pick a missing link and a nonexistent link to compare their scores, if among $n$ independent comparisons, there are $n_1$ times the missing link having a higher score and $n_2$ times have the same score, the AUC value is given as:

\begin{equation}
    \text{AUC}=\frac{n_{1} +0.5n_{2}}{n}.
\end{equation}

\textbf{(ii) AP:}  Given the ranking of the non-observed links, the precision is defined as the ratio of relevant items selected to the number of items selected. AP measures the weighted mean of precision achieved at each threshold regard the precision-recall curve, with the increase in recall used as the weight:
\begin{equation}
    \text{AP} = \sum_{n} (R_{n} - R_{n-1}) P_{n},
\end{equation}
where $R_{n}$ and $P_{n}$ are the recall and precision at the $n$th threshold.

\section{The Telecommunication Dataset}

\label{preliminaries}
 
Telecommunication networks are keys to support personal communications as well as those of businesses and other organizations in the modern era. To ensure the stability and reliability of the large telecommunication network, a crucial task is efficient fault management, which requires the precise device connections enabling the  engineers  to analyze, locate, and recover the faults, Typically, the network typologies are recovered by the Link Layer Discovery Protocol (LLDP) via analyzing the device configuration files recorded on the network management systems (NMSs).  However, as networks become increasingly complex, carriers need to maintain devices from multiple vendors and diverse standards (i.e., 2/3/4/5G), and data from different NMSs may not be associated or updated timely. It makes device connections are hard to be inferred via manually configuration file collection and protocol analysis. On the other hand, the alarm logs which describe the devices' status and latent connections are readily and easily to access. Analogous to a disease propagation network, alarms of the telecommunication network show strong temporal and spatial correlations that a fault occurring on one device triggers alarms of its own and also lead to alarms reported on connected devices. The incomplete topology also provides supportive information to infer the missing connections. 
 
 \subsection{Dataset Description}

\begin{figure}[h]
    \centering
    \includegraphics[width=0.85\linewidth]{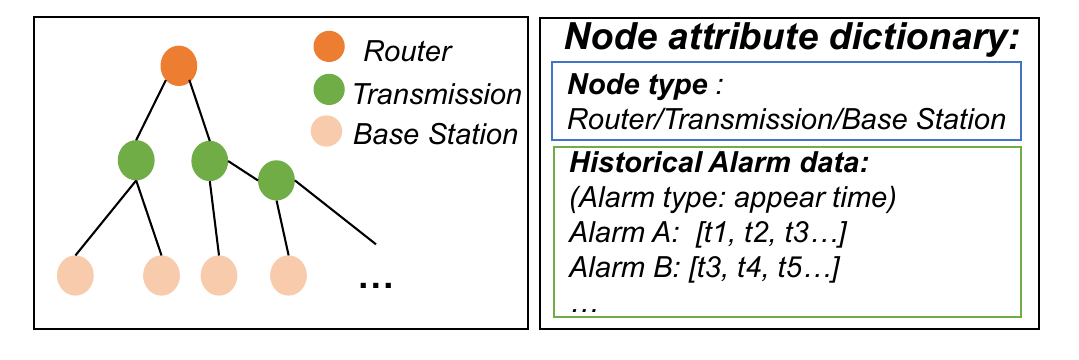}

    \caption{The  network topology (left) with alarm attributes (right) of the TeleGraph dataset.}
    \label{dataStructure}
\end{figure}

The \textit{TeleGraph} is an access layer of a metropolitan telecommunication network, which contains 41,143 vertices (devices), categorized into  3 types: routers, transmission (e.g., microwave), and base station (e.g. NodeB). The raw data includes the paths log and the alarms log, in which the paths log gives how information transits through the network. A path is an ordered list and a device may belong to multiple paths. The alarm logs tell when and where the alarms happened. Table ~\ref{table: topology information} and Table ~\ref{table:alarm data format} depict the examples of path and alarm log, respectively.  The collected alarm logs range from $12^{th}$ to $16^{th}$ April 2019, with more than six million alarms categorized into 240 types.  By combing paths and associating the alarms with the corresponding devices, we got an attributed graph in which the attributes information of each device (i.e., node) corresponds to alarm logs with each alarm type and the list of alarm occurrences of each type, sorted by time, as illustrated in Fig. ~\ref{dataStructure}. Readers may refer to the work ~\cite{fournier2020discovering} for more details on the preprocessing.

\begin{table}[!th]
\small
    \centering
\caption{Example of paths log}\begin{tabular}{c|c|c|c}
\toprule
Path Id & Device Name & Device Type & Path Hop\\
\midrule

1 & Device 1 & ROUTER &  0 \\
1 & Device 2 & MICROWAVE & 1\\ 
1 & Device 3 & NODEB  & 2\\ 
2 & Device 1 & ROUTER &  0 \\
2 & Device 4 & MICROWAVE & 1\\ 
2 & Device 5 & MICROWAVE & 2\\ 
2 & Device 6 & NODEB & 3\\ 
... & ... & ... & ... \\ 
\bottomrule
\end{tabular}
\label{table: topology information}
\vspace{-5pt}
\end{table}

\begin{table}[!th]
\small
\centering
\caption{Example of alarms log }\begin{tabular}{c|c|c|c}
\toprule
Alarm Name & Device Type & Alarm Source & Occurrence Time \\
\midrule

Alarm 1 & NODEB & Device 3 & 2019-04-12 10:40:23 \\
Alarm 2 &  MICROWAVE & Device 2& 2019-04-12 10:40:24 \\ 
Alarm 3 &   MICROWAVE & Device 2 & 2019-04-12 10:40:26 \\ 
Alarm 4 & ROUTE & Device 1 & 2019-04-12 10:40:51 \\ 
... & ... & ... & ... \\ 
\bottomrule
\end{tabular}
\label{table:alarm data format}
\vspace{-5pt}
\end{table}

 \begin{figure}[ht]
    \centering
 
    \includegraphics[width=0.75\linewidth]{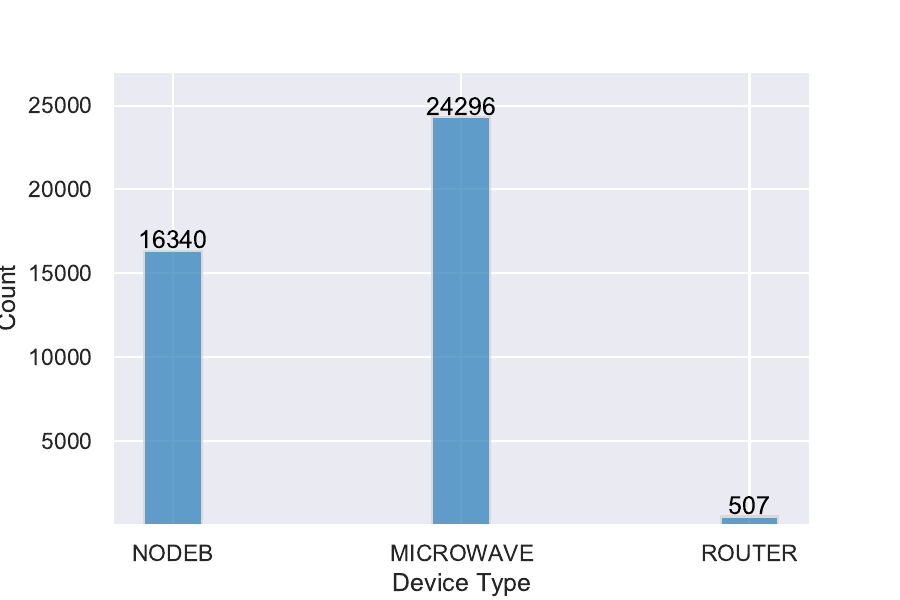}
    \caption{Number of different device types}
    \label{ne_type}
    \vspace{5pt}
\end{figure}

We first analyze the number of each type of device, which is given in Figure \ref{ne_type}. As we can see, the transmission device(i.e.,MICROWAVE) and base station (i.e., NODEB) account for most and the hub device(i.e. router) has a much small number. We further investigate the edges, which are summarized in Table \ref{tab:edge_table}. In general, a telecommunication network is bi-directed. In the downlink mode(e.g. a user receives the information from the website), the data comes from the core network flowing from routers to the base station(to terminal devices)\footnote{For the up-link mode, the information flows from the base station to microwave and then router to core network} via the microwaves, which makes the network a tree-like structure. However, the MICROWAVE may connect to the same type on the other paths, leading to the network in some areas grid or circular.

\begin{table}[!th]
\small
    \centering
    \caption{Number of edges among different devices}
    \begin{tabular}{c|c|c}
        \hline
        Source device & Target device & Number of Edges \\
        \hline
        ROUTER & ROUTER & 0 \\
        ROUTER & MICROWAVE & 2,001 \\
        ROUTER & NODEB & 22 \\
        MICROWAVE & MICROWAVE & 23,282 \\
        MICROWAVE & NODEB & 16,119 \\ 
        NODEB & NODEB & 0 \\
        \hline
        \multicolumn{2}{c|}{Total} & 41,424 \\
        \hline
    \end{tabular}
    \label{tab:edge_table}
    \vspace{-10pt}
\end{table}

 \subsection{Exploratory Analysis}

\begin{figure}[ht]
    \centering
 
    \includegraphics[width=0.85\linewidth]{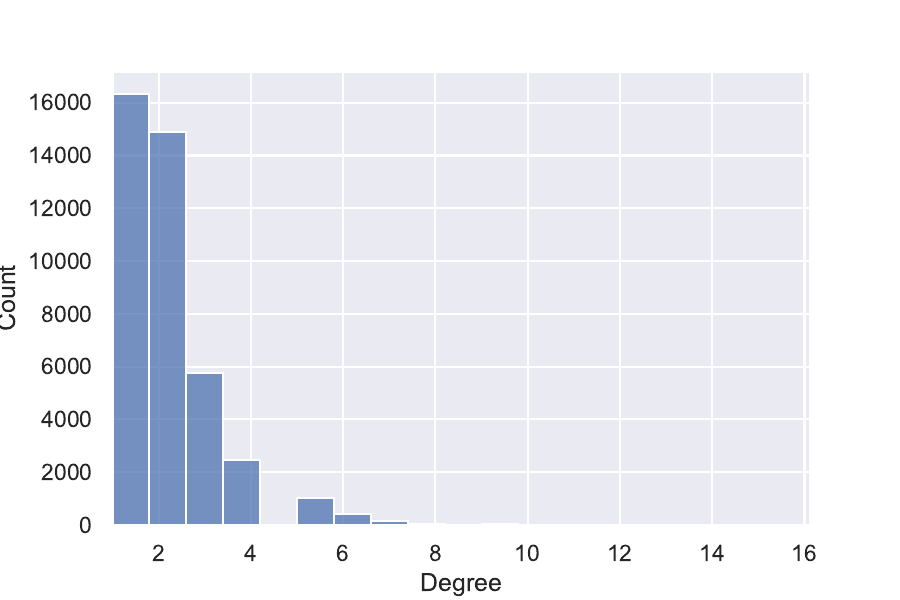}
    \caption{Degree distribution of TeleGraph, which is asymptotically power-law distributed}
    \label{degree}
    \vspace{-10pt}
\end{figure}

\begin{figure}[ht]
    \centering
 
    \includegraphics[width=0.85\linewidth]{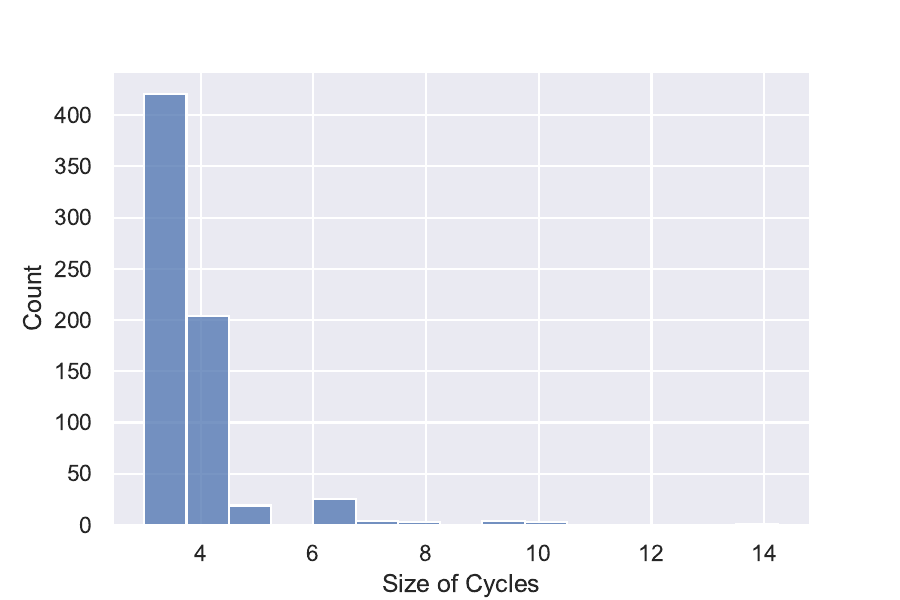}
    \caption{Node number distribution of the cycles}
    \label{fig:loop_size}
    \vspace{5pt}
\end{figure}

We further present some characteristics on the topology  of \textit{TeleGraph} and the statistics are summarized in Table ~\ref{tab:stats}, where density ~\cite{yang2022hrcf} and hyperbolicity ~\cite{chami2019hyperbolic} measure the sparsity and “tree-likeness”  of a given network, respectively.  Specifically, the density is 0 for a graph without edges and 1 for a complete graph, and the hyperbolicity value of approximately zero means a high tree-likeness. As anticipated, \textit{TeleGraph}, which has 41,143 nodes and 41,424 edges, is highly tree-like and sparse as it has very small values on both density and hyperbolicity.  We then visualize the degree distribution, as given in Figure ~\ref{degree}.  
As expected, the power-law phenomenon is very obvious with most of the nodes having a degree less than three and a very few hub nodes(i.e. routers) with distinctly large degree values.   
Though the dataset is nearly a  tree, a few loops exist with the largest ones with 13 nodes and the smallest one having only 3 nodes. We further analyze the distribution of the loop size which is further given in Figure \ref{fig:loop_size}. As observed, most of the cycles involve only 3 or 4 nodes and a small proportion of the cycles have more than 5 nodes. 

\begin{table}[h]

\small
    \centering
    \caption{Statistics of TeleGraph}
    \begin{threeparttable}
    \begin{tabular}{c|ccccc}
        \hline
     Statistics&\#Nodes&\#Edges& \#Cycles & Density $\rho$ & Hyperbolicity $\delta$\\
     \hline
     Value & 41143&41424& 684& 0.000049 &0\\
      \hline
    \end{tabular}
   
   \begin{tablenotes}
        \small
        \item[*] The smaller $\rho$ means the dataset is more sparse.
        \item[$\star$] The smaller $\delta$  indicates the dataset has a more evident tree-like structure.
    \end{tablenotes}
\end{threeparttable}
\label{tab:stats}
\vspace{-5pt}
\end{table}

In summary, the \textit{TeleGraph} is  a highly sparse and hierarchical network with rich node attributes, which is then a suitable benchmark to assess and foster link prediction techniques. However, inferring the missing links in a highly sparse and tree-like graph is challenging.  As mentioned in work ~\cite{shang2019link}, most of the widely used heuristics measures overly count on the closed triangle structures  and perform poorly in a tree-like or highly sparse network. On the other hand, most of the GNN-based link prediction algorithms are feature-centric, and how to utilize and encode the node attributes is also not easy.   

\section{Benchmark experiments}
\label{sec:exp}
In this section, we present the benchmark experiments and discuss the observations. 

\textbf{Baselines. } To investigate the link prediction algorithms on the tree-like \textit{TeleGraph} dataset, we experiment with diverse well-known baselines, including heuristic methods: common neighbors (CN)~\cite{liben2007link}, Adamic-Adar (AA)~\cite{liben2007link}, Personalized PageRank (PPR)~\cite{wang2020personalized}; embedding-based models: \textit{node2vec} ~\cite{node2vec-kdd2016}; GAE-based models ~\cite{kipf2016variational} ;
and the state-of-the-art link prediction models: NeoGNN ~\cite{yun2021neo} and SEAL~\cite{zhang2021labeling}.  Specifically, heuristic methods formulate  handcrafted rules as scoring functions.
\textit{node2vec}~\cite{node2vec-kdd2016} learns representations on graphs via optimizing a neighborhood preserving objective. For GAE-like baselines, we use different encoders including well-known Euclidean GNN models  GCN~\cite{kipf2017semi},  GAT \cite{velickovic2018graph}, and pioneering hyperbolic graph neural network HGCN~\cite{chami2019hyperbolic} \footnote{ \textit{Inner product} is employed as a decoder to compute the link existence.}. Their differences are mainly in the mechanism of message passing where GCN aggregates information through structure information while GAT aggregates information through feature correlation. 
Hyperbolic deep models generalize (graph) neural networks into hyperbolic space, abstracting underlying hierarchical layout in the graph datasets, achieving competing results of tree-like datasets on various tasks. NeoGNN~\cite{yun2021neo} learns the structural embeddings by the overlapped neighborhoods which are further combined with the node representations obtained from the feature-based GNN for link prediction.
SEAL~\cite{zhang2021labeling}  has achieved various SOTA results on link prediction tasks whose priority  mainly lies in the extraction and encoding of local structural information.

\textbf{Hyperparameter settings. } We  split the edges to 85\%/5\%/10\% for training, validation and test, respectively. And the random seed is set as 2 in data spilt. The embedding dimensions are set as 32 as well as the batch size for all models in order to make comparison fair.  
Each experiment is conducted for 10 times and we report both the mean and standard deviation. 
All the code is implemented with PyTorch and we use the implementation in PyTorch Geometric \cite{Fey/Lenssen/2019} for the GNN-related algorithms. 

\subsection{Experiment Results}

The \textit{TeleGraph} dataset provides both the device's connections as well as the non-vectorized node attributes (i.e., alarm occurrence). Graph neural networks are essentially feature-centric, which encourages us to first investigate some preliminary feature engineering schemes. Then we provide more comprehensive comparisons regarding the benchmark methods. 
\begin{table}[h]
    \centering
    \caption{Results on validation (denoted as Val) and test set of different feature schema }
    \resizebox{0.48\textwidth}{!}{
    \begin{tabular}{c|c|c c c c}\toprule
     Encoder & Embeddings & Val AUC & Val AP & Test AUC & Test AP \\
     \midrule
    \multirow{3}*{GCN} & one-hot &  69.01 $\pm$ 0.97 & 69.79 $\pm$ 0.89 & 68.74 $\pm$ 0.77 & 69.62 $\pm$ 0.64 \\
     ~ & count & 64.52 $\pm$ 0.29 & 65.38 $\pm$ 0.32 & 64.20 $\pm$ 0.30 & 65.38 $\pm$ 0.32 \\
     ~ & random & 51.87 $\pm$ 0.65 & 54.79 $\pm$ 0.69 & 51.87 $\pm$ 0.73 & 54.88 $\pm$ 0.72 \\
     \bottomrule
    \end{tabular}
    }
    \label{tab:feature}
 
\end{table}

\begin{figure}[h]
    \centering
    \includegraphics[width=0.75\linewidth]{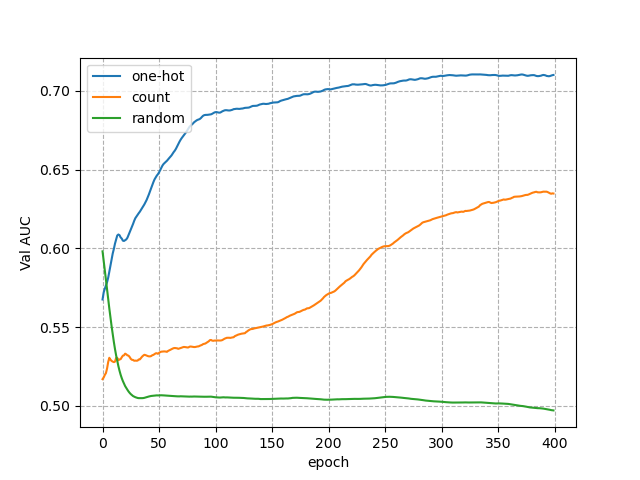}
    \caption{AUC vs Epoch on validation set}
    \label{auc_gcn}
\end{figure}

\subsubsection{Comparison on different feature schemes}
In this work, we come up with three initiatory embedding initialization schemes. In particular, \textit{one-hot} and \textit{count} assign a $1\times 240$ dimensional vector to each node according to whether the alarm appears or not and the number of occurrences, respectively. \textit{random} initializes an input embedding for each node randomly, which is widely adopted in situations where the node attributes are not available. We here  employ it to justify the informativeness of the alarm logs.  The results are summarized in Table~\ref{tab:feature} \footnote{We here only list the results of GCN due to the page limitation, both the results and analysis are consistent to that of GAT and HGCN.}.  As we can see, \textit{one-hot} produces the best performance with the AUC and AP on the test set both exceeding the second-best \textit{count} by 5$\%$. One the other hand, randomly initiating the features produces the  most unsatisfactory results that both the AUC and AP are around 0.5, which is very similar to what chance would predict. As further observed from Figure ~\ref{auc_gcn}, the AUC value of  \textit{random} even decreases with the training epoch. This is probably because the dataset is very sparse in nature, masking the validation and test edges make the message propagation and aggregation even harder and randomly assigning the features may bring some noise that hinders the learning of models.

\begin{table}[!htp]
    \small
    \centering
    \caption{ Performance Comparison on \textit{TeleGraph}.  Average values with the standard deviations are reported. For both AUC and AP, the higher, the better.}
    \resizebox{0.48\textwidth}{!}{
    \begin{tabular}{l|c c c c c c}\toprule
    \hline
    Method &  Val AUC & Val AP & Test AUC & Test AP \\ \midrule
    CN  & 51.01$\pm$ 0.00 & 51.00$\pm$ 0.00 & 51.06$\pm$ 0.00 & 51.06$\pm$ 0.00 \\
    AA  & 51.01$\pm$ 0.00 & 50.99$\pm$ 0.00 & 51.06$\pm$ 0.00 & 51.06$\pm$ 0.00 \\
    PPR & 51.76$\pm$ 0.00 & 51.60$\pm$ 0.00 & 51.75$\pm$ 0.00 & 51.83$\pm$ 0.00 \\
    Node2Vec  & 52.22 $\pm$ 0.68 & 56.34 $\pm$ 0.69 & 51.94 $\pm$ 0.51 & 56.01 $\pm$ 0.54 \\
    \hline
    GCN   &  69.01 $\pm$ 0.97 & 69.79 $\pm$ 0.89 & 68.74 $\pm$ 0.77 & 69.62 $\pm$ 0.64 \\
    GAT  & 66.80 $\pm$ 0.12 & 67.78 $\pm$ 0.13 & 67.51 $\pm$ 0.12 & 67.07 $\pm$ 0.16 \\
    HGCN & 67.49 $\pm$ 0.36 & 67.65 $\pm$ 0.24 & 67.46 $\pm$ 0.38 & 67.84 $\pm$ 0.27 \\
    NeoGNN & 63.20 $\pm$ 1.00 & 67.38 $\pm$ 0.66 & 62.45 $\pm$ 1.13 & 67.07 $\pm$ 0.70 \\
    \hline
 
    SEAL& 80.74 $\pm$ 0.15 & 80.43 $\pm$ 0.23 & 79.48 $\pm$ 0.14 & 78.82 $\pm$ 0.20 \\
    \textbf{BSAL}& 85.43 $\pm$ 0.42 & 84.21 $\pm$ 0.25 & 85.48 $\pm$ 0.21 & 84.58 $\pm$ 0.35 \\

    \midrule
 
    \bottomrule

    \end{tabular}
    }
    \label{res_huawei}
    \vspace{5pt}
\end{table}

\subsubsection{Performance comparison of baseline methods}
The experimental results of baselines methods on \textit{TeleGraph} are further summarized in Table \ref{res_huawei}\footnote{  \textit{one-hot} is adopted as the initial embedding schema for message-passing based models(i.e.,GAT, GCN, and HGCN). }.   
As anticipated, the heuristic methods (i.e., CN, AA, and PPR) which are largely based on the closed triangular structures
fail to correctly predict the link existence of the given tree-like dataset.  The AUC and AP values on both validation and test are around 0.5, which is very similar to the random guess. One the other hand, the GAE-based methods (i.e., GCN,  GAT, and HGCN) and sub-graph classification based GNN model SEAL outperform either the heuristic measures or the graph embedding model (i.e.,Node2Vec) by large margins, confirming the GNN-based solutions which formulate the link prediction task as a supervised learning problem are promising. It is noted, HGCN, which has been widely proved to be promising for tree-like dataset, is neck and neck with its Euclidean counterpart GAT. The possible reason is the handcrafted node features fail to  well-preserve the information carried on the node attributes, which hinders the representation power of hyperbolic geometry. Though NeoGNN suggests explicitly incorporating the structural information to the feature-GNNs, it still falls into the GAE framework and barely produces satisfactory performance on the given dataset. 
Last but not least, SEAL, which formulates the link prediction task as a sub-graph classification problem and further explicitly encodes the node position, shows apparent advantages over the others on this highly hierarchical and sparse  dataset. 
In summary, both the structural information and informative attributes are for link prediction. Most of the handcrafted heuristic measures heavily count on the triangle structure or preferential attachment assumption and fail to work in the highly tree-like and sparse dataset. Though the subgraph extraction and encoding scheme(i.e., SEAL) is promising, explicitly encoding and incorporating the semantic information is needed for further performance improvement.

\section{Conclusion}

Link prediction is the problem of detecting the presence of a connection between two entities in a network. Research fields, ranging from network science to machine learning and data mining, have taken a great interest in link prediction tasks. Given that hierarchical patterns are found in many real-world applications while the corresponding research datasets are inadequate, in this work, we present a new real-world dataset \textit{TeleGraph}, which is a medium-sized telecommunication network with a rich set of attributes. Our descriptive analysis of the dataset has demonstrated it is highly hierarchical and sparse, which makes the heuristic measures fail to work. We verified this precognition by a series of experiments. Our findings show that most of the available algorithms fail to produce satisfactory performance on this tree-like dataset except the subgraph-based GNN-models. More specifically, the results of a series of heuristic measures are even close to random guesses, which calls for special attention in practice. We also believe that \textit{TeleGraph} can serve as an important benchmark to assess and foster novel link prediction techniques. 

\section*{Acknowledgements}
The authors would like to thank the anonymous reviewers for their constructive suggestions.

\bibliographystyle{ACM-Reference-Format}
\bibliography{telegraph}

\appendix

\end{document}